\documentclass[final,authoryear,3p,times,twocolumn]{elsarticle}

\usepackage{amssymb}

\journal{Applied Radiation and Isotopes}

\begin{document}

\begin{frontmatter}



\title{Cross sections of deuteron induced reactions on $^{nat}$Sm for production of the therapeutic radionuclide $^{145}$Sm and $^{153}$Sm}


\author[1]{F. T\'ark\'anyi}
\author[1]{S. Tak\'acs}
\author[2]{A. Hermanne}
\author[1]{F. Ditr\'oi\corref{*}}
\author[1]{J. Csikai}
\author[3]{A.V. Ignatyuk}
\cortext[*]{Corresponding author: ditroi@atomki.hu}

\address[1]{Institute for Nuclear Research, Hungarian Academy of Sciences (ATOMKI),  Debrecen, Hungary}
\address[2]{Cyclotron Laboratory, Vrije Universiteit Brussel (VUB), Brussels, Belgium}
\address[3]{Institute of Physics and Power Engineering (IPPE), Obninsk, Russia}

\begin{abstract}
At present, targeted radiotherapy (TR) is acknowledged to have great potential in oncology. A large list of interesting radionuclides is identified, including several radioisotopes of lanthanides, amongst them $^{145}$Sm and $^{153}$Sm. In this work the possibility of their production at a cyclotron was investigated using a deuteron beam and a samarium target. The excitation functions of the $^{nat}$Sm(d,x)$^{145,153}$Sm reactions were determined for deuteron energies up to 50 MeV using the stacked-foil technique and high-resolution $\gamma$-ray spectrometry.  The measured cross sections and the contributing reactions were analyzed by comparison with results of the ALICE, EMPIRE and TALYS nuclear reaction codes. A short overview and comparison of possible production routes is given.	
\end{abstract}


\end{frontmatter}


\section{Introduction}
\label{1}
A large number of potential therapeutic radionuclides that emit low energy (conversion and 
Auger), intermediate and high energy electrons ($β^–$-emitters), or $\alpha$-particles are known \citep{Beyer,Neves,Qaim,Rosch,UusijarviMedPhys,UusijarviJNM,Zalutsky}. Among these, the presently investigated $^{145}$Sm is used for brachytherapy \citep{Fairchild,Gowda,Meigooni} and $^{153}$Sm for treatment of bone pain \citep{Bauman,Finlay,Ramamoorthy}. Both above radioisotopes can be produced through neutron capture on samarium enriched in $^{144}$Sm and $^{152}$Sm respectively \citep{IAEAReactor,IAEATherapeutic}. The production yields are high, but the products are carrier added with moderate specific activity. There is an effort to fulfill the requirements of nuclear medicine in medical radioisotope production without relying on nuclear reactors or at least without highly enriched uranium \citep{Laka,Updegraff}. The radio-lanthanides $^{145}$Sm and $^{153}$Sm can also be produced at accelerators by using charged particle or photonuclear reactions. In the frame of our systematic investigation of deuteron induced reactions on lanthanides we have measured the activation cross sections of many long-lived radio-products induced in samarium targets, among them $^{145}$Sm and $^{153}$Sm. No earlier experimental data for their excitation functions were found in the literature.
The results for production of $^{145}$Sm and $^{153}$Sm via deuteron induced reactions are compared. A discussion of the different production routes is also presented. Activation cross sections of the other investigated radioisotopes in $^{nat}$Sm(d,x) reactions will be published separately.

\section{Experiment and data evaluation}
\label{2}
The activation cross sections of the studied radioisotopes ($^{145}$Eu, $^{145}$Sm, $^{153}$Sm) were measured, relative to the $^{27}$Al(d,x)$^{22,24}$Na  and $^{nat}$Ti(d,x)$^{48}$V  monitor reactions, by using a stacked foil activation technique and $\gamma$-ray spectrometry. More details of the experiment and the data evaluations are presented in Table 1, while the decay data used are shown in Table 2. The uncertainty on each cross-section point was determined in a standard way \citep{Error} by taking the square root of the sum of the square of all relative individual contributions (except the nonlinear time parameters), supposing equal sensitivities of the different parameters appearing in the formula. The final uncertainties of the cross-sections contain uncertainties of the beam current measurement (7 \%), the number of target nuclei (5 \%), the determination of activities and conversion to absolute number of the produced nuclei (1-15 \%). The absolute values of the cross- sections are estimated to be accurate within 13 \%. 
The uncertainty of the energy scale was estimated by taking into account the energy uncertainty of the primary beam (0.3 MeV), the possible variation in the target thickness and the effect of beam straggling. The energy uncertainty in the last foil was estimated to be 1.3 MeV.

\begin{table*}[t]
\tiny
\caption{Main experimental parameters and methods of data evaluation}
\centering
\begin{center}
\begin{tabular}{|p{1.1in}|p{1.3in}|p{0.9in}|p{1.2in}|} \hline 
\textbf{Parameter} & \textbf{Value} & \textbf{Data evaluation} & \textbf{Method} \\ \hline 
Reaction  & ${}^{nat}$Sm(d,x) & Gamma spectra evaluation & Genie 2000 \citep{Canberra, Forgamma} \\ \hline 
Incident particle & Deuteron  & Determination of beam intensity & Faraday cup (preliminary)\newline Fitted monitor reaction (final) \citep{TF1991}) \\ \hline 
Method  & Stacked foil & Decay data & NUDAT 2.6 (Kinsey et al., 1997),(NuDat, 2011) \\ \hline 
Stack composition & Ho-Sm-Al-Ti, repeated 27 times & Reaction Q-values & Q-value calculator \citep{Pritychenko2003,Audi} \\ \hline 
Target and thickness  & ${}^{nat}$Sm foils, 25.14, 23.13,\newline 24.20 mm & Determination of  beam energy & Andersen (preliminary) \citep{Andersen}\newline Fitted monitor reaction (final) \citep{Ramamoorthy}  \\ \hline 
Number of target foils\newline Isotopic abundance & 27\newline 144-3.1 \%-stable\newline 147-15.0 \%-1.06*10${}^{11}$ y 148-11.3 \%-7*10${}^{15}$ y \newline 149-13.8 \%-7*10${}^{15}$ y\newline 150-7.4 \%- stable\newline 152-26.7- stable\newline 154-22.7- stable & Uncertainty of energy & cumulative effects of possible uncertainties \\ \hline 
Accelerator & Cyclone 90 cyclotron of the Université Catholique in Louvain la Neuve (LLN)  & Cross sections & Elemental cross section \\ \hline 
Primary energy & 50 MeV & Uncertainty of cross sections & sum in quadrature of all individual contributions \citep{Error} \\ \hline 
Irradiation time & 66 min & Yield & Physical yield \citep{Bonardi} \\ \hline 
Beam current & 120 nA &  &  \\ \hline 
Monitor reaction, [recommended values]  & ${}^{27}$Al(d,x)${}^{24}$Na \newline ${}^{nat}$Ti(d,x)${}^{48}$V reactions &  & \citep{TF2001} \\ \hline 
Monitor target and thickness & ${}^{nat}$Al, 98, 49.6 mm\newline ${}^{nat}$Ti  10.9 mm &  &  \\ \hline 
detector & HPGe &  &  \\ \hline 
g-spectra measurements & 4 series &  &  \\ \hline 
Cooling times & 6 h, 25 h, 70 h, 530 h &  &  \\ \hline 
\end{tabular}

\end{center}
\end{table*}

\begin{table*}[t]
\tiny
\caption{Decay characteristics of the investigated activation products and Q-values of contributing reactions}
\centering
\begin{center}
\begin{tabular}{|p{0.7in}|p{0.6in}|p{0.7in}|p{0.5in}|p{0.8in}|p{0.8in}|} \hline 
Nuclide\newline Decay path & Half-life & E$_{\gamma}$(keV) & I$_{\gamma}$(\%) & Contributing reaction & Q-value\newline (keV) \\ \hline 
\textbf{${}^{1}$${}^{45}$Eu\newline }~$\varepsilon $: 100 \%~\textbf{} & 5.93 d & ~653.512\newline ~893.73\newline ~1658.53 & 15.0 \newline 66 \newline 14.9  & ${}^{144}$Sm(d,n)\newline ${}^{14}$${}^{7}$Sm(d,4n)\newline ${}^{14}$${}^{8}$Sm(d,5n)\newline ${}^{14}$${}^{9}$Sm(d,6n)\newline ${}^{150}$Sm(d,7n)\newline ${}^{1}$${}^{52}$Sm(d,9n)\newline ${}^{1}$${}^{54}$Sm(d,11n) & 1090.5~\newline -20424.02\newline ~-28565.4\newline ~-34435.74~\newline ~-42422.43\newline ~-56276.58 \\ \hline 
\textbf{${}^{1}$${}^{45}$Sm\newline }$\varepsilon $: 100 \%~\textbf{} & 340 d & 61.2265 & ~12.15  & ${}^{144}$Sm(d,p)\newline ${}^{14}$${}^{7}$Sm(d,p3n)\newline ${}^{14}$${}^{8}$Sm(d,p4n)\newline ${}^{14}$${}^{9}$Sm(d,p5n)\newline ${}^{150}$Sm(d,p6n)\newline ${}^{1}$${}^{52}$Sm(d,p8n)\newline ${}^{145}$Eu decay\newline ${}^{145}$Pm decay & 4532.534\newline -16981.99\newline -25123.37\newline -30993.72\newline -38980.41\newline -52834.55~\newline 1090.5\newline ~-15583.55 \\ \hline 
\textbf{${}^{1}$${}^{53}$Sm\newline }$\beta $${}^{-}$: 100 \%\textbf{} & 46.28 h & 103.18012\newline  & 29.25  & ${}^{1}$${}^{52}$Sm(d,p)\newline ${}^{1}$${}^{54}$Sm(d,p2n)\newline ${}^{153}$Pm decay & 3643.834\newline ~~-10191.36\newline -11320.47 \\ \hline 
\end{tabular}

\end{center}
\begin{flushleft}
\footnotesize{\noindent When complex particles are emitted instead of individual protons and neutrons the Q-values have to be decreased by the respective binding energies of the compound particles: np-d, +2.2 MeV; 2np-t, +8.48 MeV; 

\noindent n2p-${}^{3}$He, +7.72 MeV; 2n2p-a, +28.30 MeV}

\end{flushleft}

\end{table*}

\begin{figure}
\includegraphics[width=0.5\textwidth]{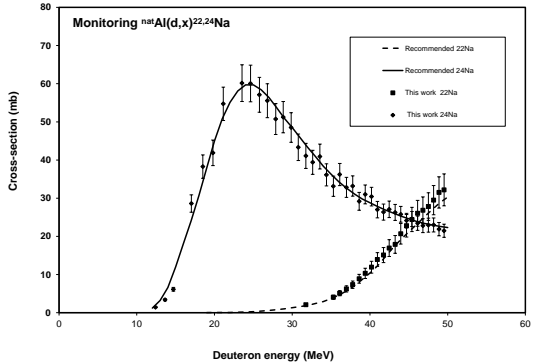}
\caption{The simultaneously re-measured monitor reactions }
\label{fig:1}       
\end{figure}

\section{Model calculations}
\label{3}
The updated ALICE-IPPE \citep{Dityuk} and EMPIRE \citep{Herman} codes (the modified versions for deuteron induced reactions ALICE-IPPE-D and EMPIRE-D) were used to analyse the present experimental results. These modifications were developed and implemented in the original codes at IPPE and more details can be found in our previous reports \citep{Hermanne,TF2007}. In these codes a simulation of direct (d,p) and (d,t) transitions by the general relations for a nucleon transfer probability in the continuum is included through an energy dependent enhancement factor for the corresponding transitions. The phenomenological enhancement factor is based on systematics of experimental data of the related reactions \citep{Ignatyuk}. The theoretical data from the TENDL-2012 \citep{Koning2012} library (based on the most recent TALYS code \citep{Koning2007} were also included in the comparison.

\section{Results}
\label{4}
According to Table 1, the $^{145}$Sm (T$_{1/2}$ = 340 d) is produced directly and through the decay of $^{145}$Eu (T$_{1/2}$ = 5.93 d). Both products were identified in our spectra. Possible contribution from long-lived $^{145}$Pm (T$_{1/2}$ = 17.7 a) can be ignored in this experiment. The situation is similar for $^{153}$Sm (T$_{1/2}$ = 46.50 h), but in this case the parent $^{153}$Pm (T$_{1/2}$ = 5.3 min) was not detected due to the short half-life and the low predicted cross sections.

\subsection{Excitation functions}
\label{4.1}
The cross sections for all the reactions studied are shown in Figures 2 – 5 and the numerical values are shown in Table 3. The results for the 40 and 20 MeV primary incident energies are shown separately to point out the agreement in the overlapping energy range. The reactions responsible for the production of the given activation products and their Q-values, obtained from the calculator of the Brookhaven Nat. Lab. \citep{Pritychenko2003}, are given in Table 2.

\subsubsection{$^{145}$Eu  (direct production)}
\label{4.1.1}
The radioisotope $^{145}$Eu (T$_{1/2}$ = 5.93 d) is formed at low energies by the $^{144}$Sm(d,n) reaction  (first maximum) and by $^{147-150}$Sm(d,4-7n) reactions above 20 MeV. The measured experimental data are shown in Fig. 2, together with the theoretical predictions. The results in TENDL-2012 slightly overestimate both the low and high energy experimental data. The overestimation is more significant in the case of EMPIRE-D and ALICE-D.

\begin{figure}
\includegraphics[width=0.5\textwidth]{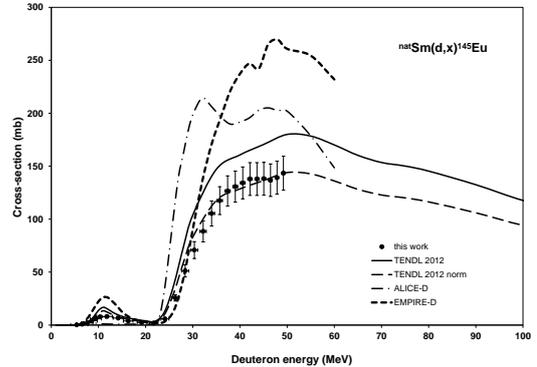}
\caption{Experimental and theoretical excitation functions for the $^{nat}$Sm(d,xn)$^{145}$Eu  reaction}
\label{fig:2}       
\end{figure}


\subsubsection{$^{145}$Sm  (cumulative)}
\label{4.1.2}
The cumulative cross sections for $^{145}$Sm (T$_{1/2}$ = 340 d) production are shown in Fig. 3.  In this experiment $^{145}$Sm is produced directly via (d,pxn) reactions and from decay of the shorter-lived $^{145}$Eu (T$_{1/2}$ = 5.93 d) discussed above. The cross sections were deduced from spectra measured after 5 half-lives of the $^{145}$Eu parent decaying completely with EC to $^{145}$Sm (97 \%). The missing part was corrected on the basis of the measured $^{145}$Eu cross sections. As stated earlier, the possible contribution from the long-lived $^{145}$Pm (T$_{1/2}$ = 17.7 a) can be ignored. The first peak of the excitation function at 10 MeV is the sum of the $^{144}$Sm(d,n) and $^{144}$Sm(d,p) reactions. The higher energy part is the sum of $^{147-150}$Sm(d,4-7n) and $^{147-150}$Sm(d,p3-6n) reactions. There is a good agreement between the cumulative production of $^{145}$Sm and the results of TENDL-2012 at higher energies. At energies below 15 MeV, as a result of the known underestimation of (d,p)  \citep{Hermanne,TF2007} and  “compensation” by overestimation of (d,n) reaction, the TENDL discrepancy is finally only a few percent. In case of ALICE-D and EMPIRE-D at low energies the description of the (d,p) is better, but the high energy part is significantly overestimated.

\begin{figure}
\includegraphics[width=0.5\textwidth]{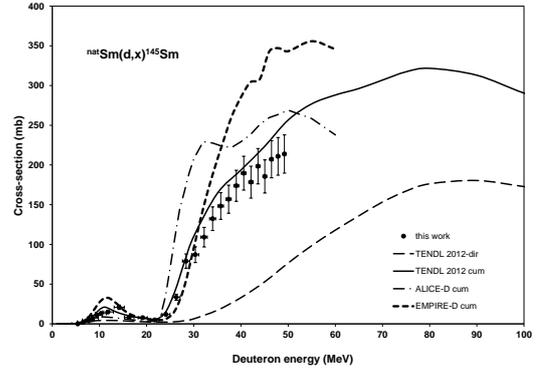}
\caption{Experimental and theoretical excitation functions for $^{nat}$Sm(d,x)$^{145}$Sm}
\label{fig:3}       
\end{figure}

\subsubsection{$^{153}$Sm}
\label{4.1.3}
The radioisotope $^{153}$Sm (T$_{1/2}$ = 46.50 h) is formed directly by the $^{152}$Sm(d,p) and $^{154}$Sm(d,p2n) reactions and from the decay of short-lived $^{153}$Pm (T$_{1/2}$ = 5.25 min). The experimental and theoretical results are shown in Fig. 4. The $^{153}$Pm is produced via the $^{154}$Sm(d,2pn) low cross section reaction. The TENDL-2012 data underestimate the (d,p) reaction part and overestimate  the (d,p2n). In case of ALICE-D and EMPIRE-D the agreement for low energy (d,p) part is good, but the (d,p2n) is underestimated.

\begin{figure}
\includegraphics[width=0.5\textwidth]{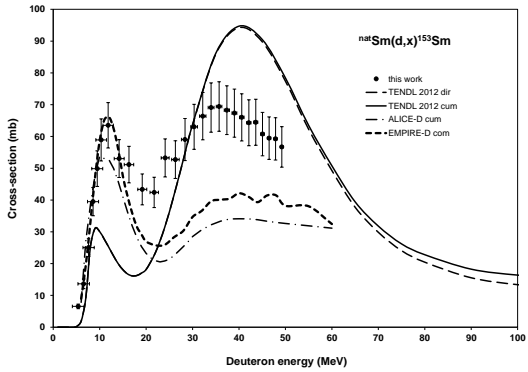}
\caption{Experimental and theoretical excitation functions for $^{nat}$Sm(d,x)$^{153}$Sm}
\label{fig:4}       
\end{figure}

\begin{table*}[t]
\tiny
\caption{Measured cross sections of the \textit{${}^{1}$${}^{4}$${}^{5}$Eu and ${}^{1}$${}^{45,153}$Sm }reactions and estimated uncertainties}
\centering
\begin{center}
\begin{tabular}{|c|c|c|c|c|c|c|c|c|c|c|c|} \hline 
\multicolumn{3}{|c|}{\textbf{Energy (MeV) }} & \multicolumn{3}{|c|}{${}^{145}$Eu\textbf{ (mb)}} & \multicolumn{3}{|c|}{${}^{145}$Sm\textbf{  (mb)}} & \multicolumn{3}{|c|}{${}^{153}$Sm\textbf{ \newline (mb)}} \\ \hline 
49.2 & $\pm$ & 0.30 & 143.4 & $\pm$ & 16.10 & 213.9 & $\pm$ & 24.1 & 56.7 & $\pm$ & 6.4 \\ \hline 
47.9 & $\pm$ & 0.33 & 139.2 & $\pm$ & 15.64 & 210.9 & $\pm$ & 23.7 & 59.2 & $\pm$ & 6.7 \\ \hline 
46.5 & $\pm$ & 0.36 & 136.8 & $\pm$ & 15.37 & 207.2 & $\pm$ & 23.3 & 59.5 & $\pm$ & 6.7 \\ \hline 
45.1 & $\pm$ & 0.39 & 138.2 & $\pm$ & 15.51 & 185.6 & $\pm$ & 20.9 & 60.8 & $\pm$ & 6.8 \\ \hline 
43.6 & $\pm$ & 0.42 & 137.9 & $\pm$ & 15.49 & 198.4 & $\pm$ & 22.3 & 64.5 & $\pm$ & 7.2 \\ \hline 
42.1 & $\pm$ & 0.46 & 137.9 & $\pm$ & 15.50 & 178.4 & $\pm$ & 20.1 & 64.3 & $\pm$ & 7.2 \\ \hline 
40.6 & $\pm$ & 0.49 & 134.2 & $\pm$ & 15.08 & 189.7 & $\pm$ & 21.4 & 66.0 & $\pm$ & 7.4 \\ \hline 
39.0 & $\pm$ & 0.52 & 130.7 & $\pm$ & 14.68 & 173.9 & $\pm$ & 19.6 & 67.3 & $\pm$ & 7.6 \\ \hline 
37.4 & $\pm$ & 0.56 & 126.6 & $\pm$ & 14.23 & 157.0 & $\pm$ & 17.7 & 68.3 & $\pm$ & 7.7 \\ \hline 
35.7 & $\pm$ & 0.60 & 117.5 & $\pm$ & 13.21 & 148.3 & $\pm$ & 16.8 & 69.4 & $\pm$ & 7.8 \\ \hline 
34.0 & $\pm$ & 0.63 & 105.4 & $\pm$ & 11.84 & 132.4 & $\pm$ & 15.0 & 69.1 & $\pm$ & 7.8 \\ \hline 
32.2 & $\pm$ & 0.67 & 88.5 & $\pm$ & 9.95 & 109.1 & $\pm$ & 12.3 & 66.4 & $\pm$ & 7.5 \\ \hline 
30.3 & $\pm$ & 0.72 & 70.8 & $\pm$ & 7.96 & 87.1 & $\pm$ & 9.9 & 63.0 & $\pm$ & 7.1 \\ \hline 
28.4 & $\pm$ & 0.76 & 51.2 & $\pm$ & 5.76 & 79.0 & $\pm$ & 9.0 & 59.0 & $\pm$ & 6.6 \\ \hline 
26.3 & $\pm$ & 0.80 & 25.6 & $\pm$ & 2.89 & 33.6 & $\pm$ & 3.9 & 52.7 & $\pm$ & 5.9 \\ \hline 
24.1 & $\pm$ & 0.85 & 5.9 & $\pm$ & 0.67 & 11.8 & $\pm$ & 1.4 & 53.3 & $\pm$ & 6.0 \\ \hline 
21.7 & $\pm$ & 0.90 & 1.9 & $\pm$ & 0.24 & 4.9 & $\pm$ & 1.4 & 42.4 & $\pm$ & 4.8 \\ \hline 
19.2 & $\pm$ & 0.96 & 2.4 & $\pm$ & 0.30 & 7.8 & $\pm$ & 1.1 & 43.4 & $\pm$ & 4.9 \\ \hline 
16.3 & $\pm$ & 1.02 & 3.9 & $\pm$ & 0.46 & 8.4 & $\pm$ & 1.0 & 51.1 & $\pm$ & 5.7 \\ \hline 
14.2 & $\pm$ & 1.07 & 6.6 & $\pm$ & 0.77 & 20.9 & $\pm$ & 2.8 & 53.0 & $\pm$ & 6.0 \\ \hline 
11.8 & $\pm$ & 1.12 & 8.2 & $\pm$ & 0.93 & 15.0 & $\pm$ & 1.7 & 63.5 & $\pm$ & 7.1 \\ \hline 
10.3 & $\pm$ & 1.16 & 7.7 & $\pm$ & 0.88 & 13.1 & $\pm$ & 1.5 & 58.9 & $\pm$ & 6.6 \\ \hline 
9.5 & $\pm$ & 1.17 & 6.0 & $\pm$ & 0.69 & 10.8 & $\pm$ & 1.3 & 49.9 & $\pm$ & 5.6 \\ \hline 
8.6 & $\pm$ & 1.19 & 4.1 & $\pm$ & 0.47 & 7.2 & $\pm$ & 0.8 & 39.5 & $\pm$ & 4.4 \\ \hline 
7.6 & $\pm$ & 1.22 & 2.0 & $\pm$ & 0.23 & 4.6 & $\pm$ & 0.5 & 25.0 & $\pm$ & 2.8 \\ \hline 
6.6 & $\pm$ & 1.24 & 0.9 & $\pm$ & 0.10 & 2.3 & $\pm$ & 0.3 & 13.6 & $\pm$ & 1.5 \\ \hline 
5.4 & $\pm$ & 1.26 & 0.3 & $\pm$ & 0.03 &  &  &  & 6.5 & $\pm$ & 0.7 \\ \hline 
\end{tabular}

\end{center}
\end{table*}

\subsection{Thick target yields }
\label{4.2}
Thick target yields (integrated yield for a given incident energy down to the reaction threshold) were calculated from curves fitted to our experimental cross section data The results for physical yields \citep{Bonardi} are presented in Fig. 5. No earlier experimental thick target yield data were found in the literature.

\begin{figure}
\includegraphics[width=0.5\textwidth]{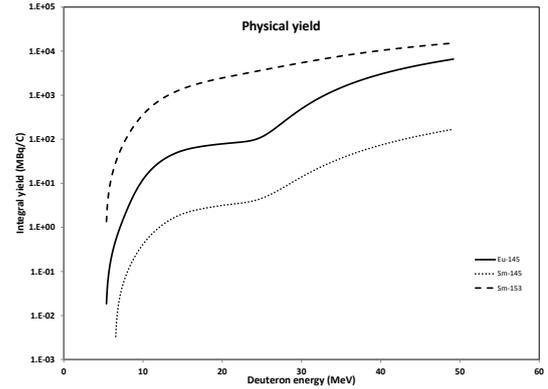}
\caption{Integral yields for production of $^{145}$Eu and $^{145,153}$Sm deduced from the excitation functions}
\label{fig:5}       
\end{figure}

\section{Comparison of the production routes}
\label{5.}
The two radio-lanthanides discussed ($^{145,153}$Sm) can be produced via various reactions. Among these competitive production routes we discuss below only the low energy, light charged particle induced routes, presented in Table 4 in more detail (E$_{particle}  <$  100 MeV). Table 4 does not contain the $^3$He induced reactions from practical reasons, taking into account the availability and cost of the high intensity $^3$He irradiations. The related excitation functions are shown in Figs. 6-11.  Except for the $^{nat}$Sm(p,xn)$^{145}$Eu-$^{145}$Sm reaction, investigated by Blue et al.  (Blue, 1988) (yields) and the present data on deuterons, no experimental data are available for cross sections and yields for charged particle production routes. The (n,$\gamma$) cross sections and yields were taken from \citep{Mirzadeh,Pritychenko2012}. For comparison of production routes of charged particle induced reactions we have used TTY (thick target yield) derived from the theoretical data of TENDL, ALICE and EMPIRE which are normalized to the measured experimental data, in case of existing experimental data. Otherwise, the theoretical data were selected after comparing with the systematics of the experimental data in the same mass region.
As for many practical applications, also for medical isotope production there are several factors determining the optimal production route. 
The first group of factors is connected to requirements in the patient related application: level of specific activity, radionuclidic purity, carrier added or carrier free product, large scale routine application, small scale research work route.
The second group of factors is connected to the technology used: minimal required production yields, natural or highly enriched targets, target mass, target cooling problems, recovery and radioactive waste problems, the available beam properties (particle type, current, energy), possibility of parallel irradiations, etc.
The third group of factors is determined by the continuously changing day to day requirements: the new competing radio-products and application methods, the new regulations and trend in nuclear technology.
Nowadays the largest part of routinely used therapeutic radioisotopes is produced at nuclear research reactors. There is however a significant effort to transfer the production to accelerators by improving beam characteristics and production technology and/or introducing new accelerator produced isotopes with characteristics similar to reactor products.

\subsection{Production of $^{145}$Sm}
\label{5.1}
According to Table 4, the $^{144}$Sm(n,$\gamma$)$^{145}$Sm production route is the method of choice from point of view of yield and irradiation technology if a carrier added, relatively low specific activity $^{145}$Sm end product is satisfactory.
The route to get carrier free, high specific activity $^{145}$Sm  is possible only with accelerators  through the decay of $^{145}$Eu or through the direct production from alpha induced nuclear reactions on natural or enriched neodymium.  Indirect production through $^{145}$Eu obtained from irradiation of Sm targets requires two chemical separations, while in the case of direct production one chemical separation is needed to produce a high-specific-activity and high radionuclidic purity $^{145}$Sm end product.
Out of them the $^{147}$Sm(p,3n)$^{145}$Eu-$^{145}$Sm and the $^{147}$Sm(p,x)$^{145}$Sm are the most productive ($^{147}$Sm 15\% in natural) and requires a 30 MeV cyclotron (or accelerator). The product is contaminated with long lived $\alpha$-emitter $^{146}$Sm, but with very low activity. The radionuclidic impurity is lower in the case of the $^{144}$Sm(d,n)$^{145}$Eu-$^{145}$Sm ($^{144}$Sm 3.1\%, low energy irradiation to avoid (d,3n) and contamination with long-lived $^{143}$Pm) and $^{142}$Nd(${\alpha}$,n)$^{145}$Sm reactions (low energy to avoid (${\alpha}$,2n) and (${\alpha}$,3n)). In case of other p-, d-, and ${\alpha}$ induced reactions the simultaneously produced $^{146,147}$Sm do not cause serious impurity due to their extremely long half-lives. 
In all cases highly enriched targets are required. The excitation functions for the main reactions based on our ALICE-IPPE, EMPIRE results and on data in TENDL-2012, are shown in Figs. 6-9. 

\begin{figure}
\includegraphics[width=0.5\textwidth]{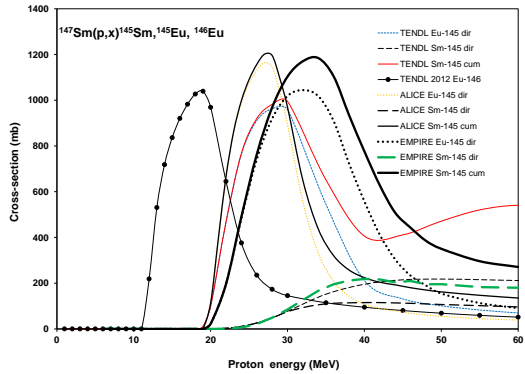}
\caption{Theoretical excitation functions for $^{147}$Sm(p,x)$^{145}$Eu and $^{145}$Sm reactions}
\label{fig:6}       
\end{figure}

\begin{figure}
\includegraphics[width=0.5\textwidth]{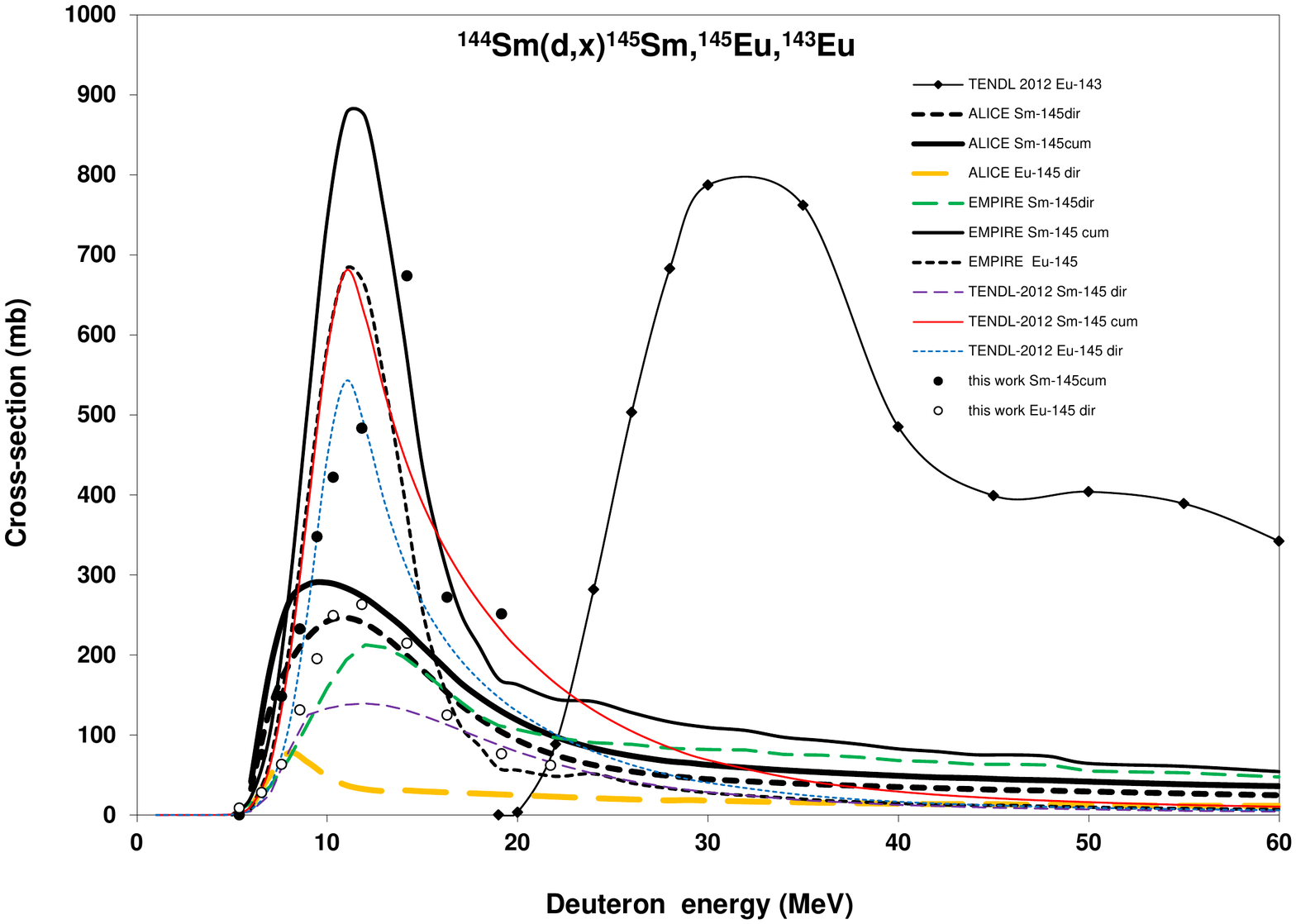}
\caption{Theoretical excitation functions for $^{147}$Sm(d,xn) $^{145}$Sm and $^{145}$Eu reactions}
\label{fig:7}       
\end{figure}

\begin{figure}
\includegraphics[width=0.5\textwidth]{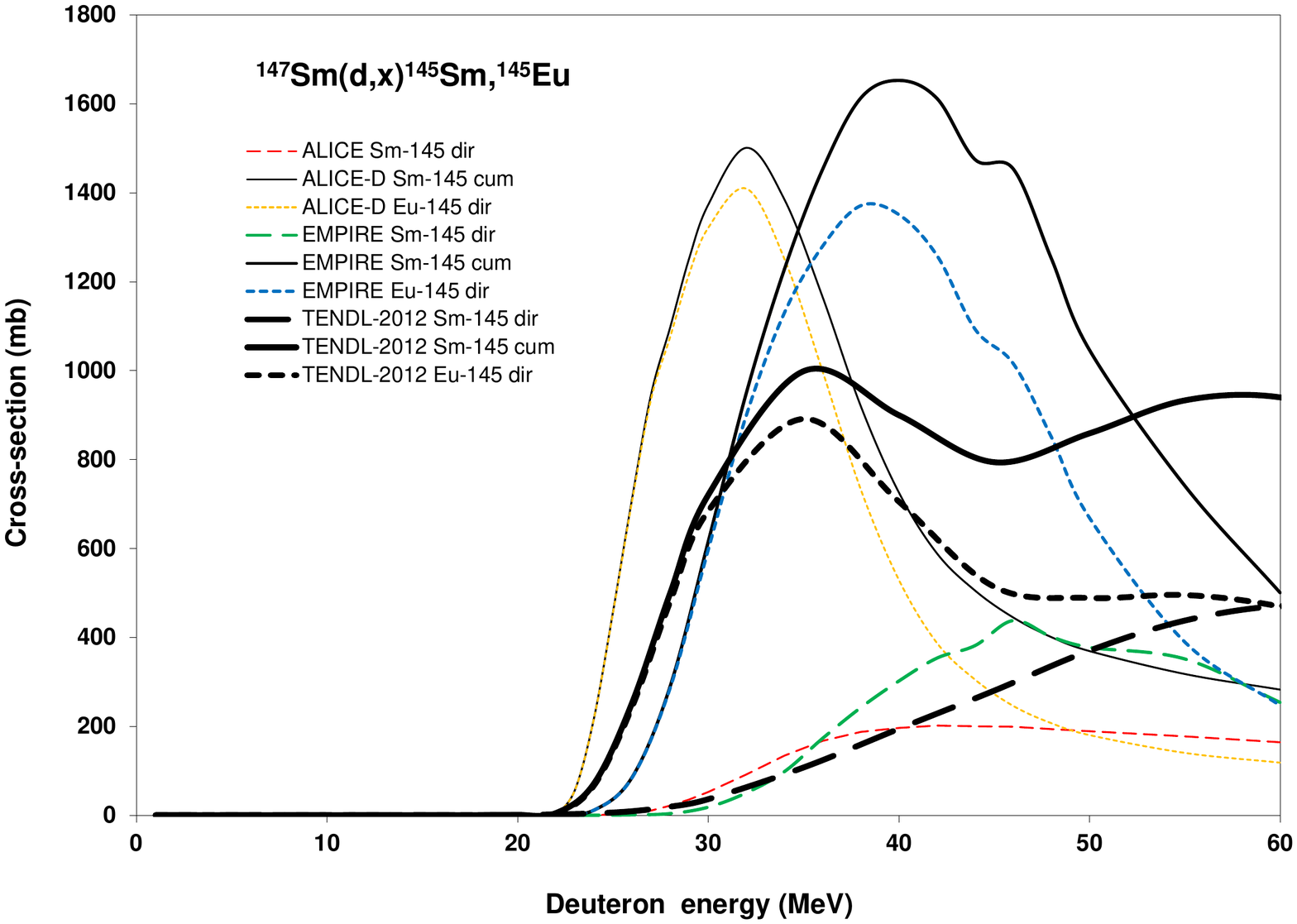}
\caption{Theoretical excitation functions for $^{147}$Sm(d,xn) $^{145}$Sm and $^{145}$Eu reactions}
\label{fig:8}       
\end{figure}

\subsection{Production of $^{153}$Sm}
\label{5.2}
According to the Table 4, it is difficult to compete in efficiency with the production of $^{153}$Sm in reactors due to the large production cross section and also because the products of the proton and deuteron induced reactions are carrier added. The only way to get a no carrier added product is through the low yield $^{150}$Nd(${\alpha}$,n)$^{153}$Sm reaction (5.1\% abundance). In all cases highly enriched targets are required.
The proton and deuteron induced reactions can however have importance for local use due to the relatively short half-life of $^{153}$Sm (T$_{1/2}$ =  46.50 h) and due to the unused beam time of charged particle accelerators. The production requires highly enriched targets. The cross sections of the most important reactions are shown in Figs. 10-12. 

\begin{figure}
\includegraphics[width=0.5\textwidth]{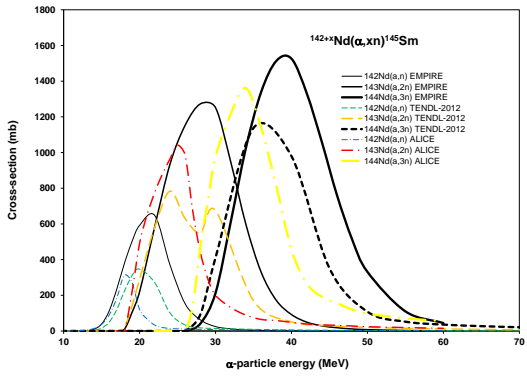}
\caption{Experimental and theoretical excitation functions for $^{142}$Nd(${\alpha}$,xn)$^{145,144,143}$Sm reactions}
\label{fig:9}       
\end{figure}

\begin{figure}
\includegraphics[width=0.5\textwidth]{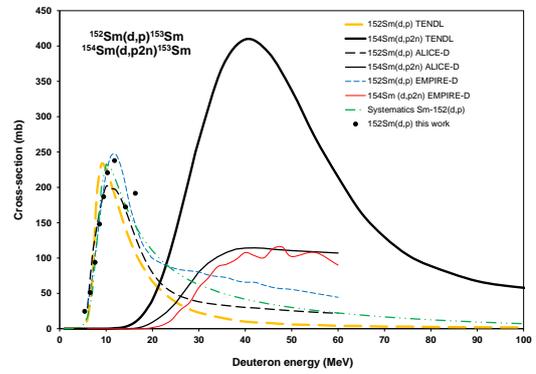}
\caption{Theoretical and experimental excitation functions for $^{152}$Sm(d,p)$^{153}$Sm  and $^{154}$Sm(d,p2n)$^{153}$Sm reactions}
\label{fig:10}       
\end{figure}

\begin{figure}
\includegraphics[width=0.5\textwidth]{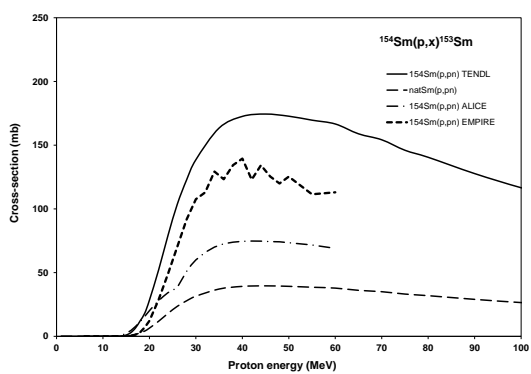}
\caption{Theoretical excitation functions for $^{154}$Sm(p,pn)$^{153}$Sm reactions}
\label{fig:11}       
\end{figure}

\begin{figure}
\includegraphics[width=0.5\textwidth]{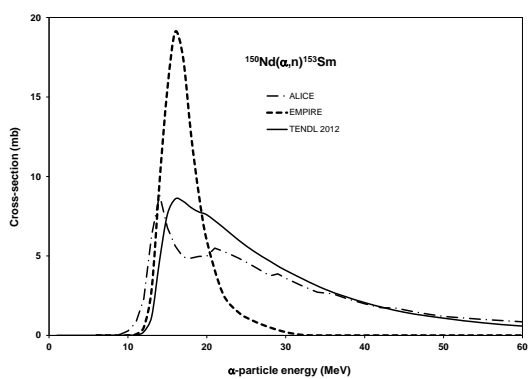}
\caption{Theoretical excitation functions for $^{142}$Nd(${\alpha}$,n)$^{153}$Sm reaction}
\label{fig:12}       
\end{figure}

\begin{table*}[t]
\tiny
\caption{Production routes of ${}^{1}$${}^{4}$${}^{5}$Sm and ${}^{15}$${}^{3}$Sm}
\centering
\begin{center}
\begin{tabular}{|c|c|c|c|c|c|} \hline 
Nuclide & Reaction & $\sigma, \sigma_{max}$\newline (b) & Energy range\newline (MeV) & Yield\newline (MBq/C) & Reference \\ \hline 
${}^{145}$Sm & ${}^{144}$Sm(n,$\gamma$)${}^{145}$Sm & 1.64  & thermal & 200 MBq/mg${}^{*}$ & \citep{Mirzadeh} \\ \hline 
${}^{145}$Sm & ${}^{144}$Sm(d,n)${}^{145}$Eu-${}^{145}$Sm & 0.26 & 20-8 & 32 & this work (exp) \\ \hline 
 & ${}^{144}$Sm(d,x)${}^{145}$Sm & 0.68 & 20-8 & 97 & this work (exp) \\ \hline 
 & ${}^{147}$Sm(d,4n)${}^{145}$Eu-${}^{145}$Sm & 0.89 & 50-25 & 713 & TENDL 2012 \\ \hline 
 & ${}^{147}$Sm(d,x)${}^{145}$Sm & 0.99 & 50-25 & 914 & TENDL 2012 \\ \hline 
 & ${}^{nat}$Sm(d,xn)${}^{145}$Eu-${}^{145}$Sm & 0.15 & 50-25 & 121 & this work (exp) \\ \hline 
 & ${}^{nat}$Sm(d,x)${}^{145}$Sm & 0.26 & 50-25 & 163 & this work (exp) \\ \hline 
 & ${}^{147}$Sm(p,3n)${}^{145}$Eu-${}^{145}$Sm & 1.05 & 50-20 & 64854 & this work(EMPIRE)  \\ \hline 
 & ${}^{147}$Sm(p,x)${}^{145}$Sm & 1.18 & 50-20 & 81906 & this work(EMPIRE)  \\ \hline 
 & ${}^{142}$Nd($\alpha$,n)${}^{145}$Sm & 0.31 & 24-14 & 289 & this work(ALICE) \\ \hline 
 & ${}^{143}$Nd(a,2n)${}^{145}$Sm & 1.05 & 35-19 & 2500 & this work(ALICE) \\ \hline 
${}^{153}$Sm & ${}^{152}$Sm(n,$\gamma$)${}^{153}$Sm & 206   & thermal & 10${}^{5}$ MBq/mg${}^{*}$${}^{,}$${}^{\#}$ & \citep{Mirzadeh} \\ \hline 
${}^{153}$Sm & ${}^{152}$Sm(d,p)${}^{153}$Sm & 0.25 & 24-7 & 10057 & this work,  (exp+ EMPIRE) \\ \hline 
 & ${}^{154}$Sm(d,p2n)${}^{153}$Sm & 0.3 & 50-20 & 54639 & this work (exp+EMPIRE) \\ \hline 
 & ${}^{154}$Sm(p,pn)${}^{153}$Sm & 0.14 & 100-20 & 145550 & this work(EMPIRE) \\ \hline 
 & ${}^{nat}$Sm(p,pxn)${}^{153}$Sm & 0.031 & 100-20 & 47480 & this work(EMPIRE)  \\ \hline 
 & ${}^{150}$Nd($\alpha$,n)${}^{ 153}$Sm & 0.020 & 40-15 & 66 & this work (EMPIRE) \\ \hline 
\end{tabular}

\begin{flushleft}
\footnotesize{\noindent ${}^{*}$ irradiation time = half-life; neutron flux = 1.10${}^{14}$${}^{ }$n/cm${}^{2}$/s; 100 \% enriched target

\noindent ${}^{\#}$${}^{ }$irradiation time = 180 d}

\end{flushleft}
\end{center}
\end{table*}

\section{Summary and conclusion}
\label{6.}
From a series of radioisotopes produced in deuteron induced nuclear reactions on natural samarium the medically important $^{145}$Sm and $^{153}$Sm as well as $^{145}$Eu as parent of $^{145}$Sm were selected for detailed study. Experimental excitation functions of both isotopes were accurately determined and compared with the results of three nuclear reaction codes. The agreement with the theoretical results and the present experiments was discussed for each isotope separately, because the different codes gave different quality estimations in the case of the above three radioisotopes. From the measured excitation functions yield curves were also calculated in the investigated energy range. Since both samarium isotopes are important for medical applications, a comparison was also made between all possible routes. It turned out, as expected, that the neutron induced production in reactors is the most economical way for mass production of these isotope in carrier added form. However, it should be mentioned that for no carrier added end product and small scale or research production charged particle induced reactions can be used as an alternative.

\section{Acknowledgements}
\label{}
This work was done in the frame of MTA-FWO research project. The authors acknowledge the support of research projects and of their respective institutions in providing the materials and the facilities for this work.
 



\bibliographystyle{elsarticle-harv}
\bibliography{Smd}







\end{document}